\documentclass[%
 reprint,
 amsmath,amssymb,
 aps,
 prl,
 superscriptaddress
]{revtex4-2}
\usepackage{graphicx}
\usepackage{dcolumn}
\usepackage{bm}
\usepackage{xcolor}

\begin{document}

\title{Directional Photocurrent Generated by Quantum Interference Control}

\author{Yiming Gong}
\affiliation{Department of Physics, University of Michigan, Ann Arbor, MI, 48109, USA}

\author{Kai Wang}
\affiliation{Department of Physics, University of Michigan, Ann Arbor, MI, 48109, USA}
\affiliation{School of Physics and Astronomy, Zhuhai Campus,Sun Yat-sen University, Zhuhai, GD, 519082, China}

\author{Steven T. Cundiff}
\email{cundiff@umich.edu}
\affiliation{Department of Physics, University of Michigan, Ann Arbor, MI, 48109, USA}
\affiliation{Quantum Research Institute, University of Michigan, Ann Arbor, MI 48109 USA}

\date{\today}

\begin{abstract}
Although the absorption of light in a bulk homogeneous semiconductor produces photocarriers with non-zero momentum, it generally does not produce a current in the absence of an applied electric field because equal amounts of carriers with opposite momentum are injected. The interference of absorption processes, for example, between one-photon and two-photon absorption, can produce a current because constructive interference for carriers with one momentum can correspond to destructive interference for carriers with the opposite momentum. We show that for the interference between two-photon and three-photon absorption, the current has a narrower angular spread, i.e., a ``beam'' of electrons in a specified direction is produced in the semiconductor.
\end{abstract}

\maketitle


Interference is one of the hallmarks of a quantum process; it simply does not occur for classical particles. Quantum interference occurs when the wavefunction of a particle can traverse two, or more, separate pathways to a common final state. The relative phase accumulated along the pathways determines if the interference is constructive or destructive. This phase dependence was recognized to provide a method, which does not exist classically, of controlling processes~\cite{Brumer1986,Shapiro2012}. One realization of this method is quantum interference control (QuIC) between optical processes with differing numbers of photons that excite a target state. QuIC has been used to control the direction of atomic ionization~\cite{Chen1990,Yin1992,Eickhoff2021,Sun2021,Ohmura2025} and molecular ionization and disassociation~\cite{Brumer1986,Shapiro2012,Nagai2006,Sheehy1995,Zhu1995}. In these experiments, a two-color field, where the frequencies of the two colors are integer related, was used. Since a two-color field can have a phase-dependent broken inversion symmetry, it is not surprising that it can result in ionization or disassociation that has broken inversion symmetry and is phase dependent.

\begin{figure}[t]
\includegraphics[width=1.0\columnwidth]{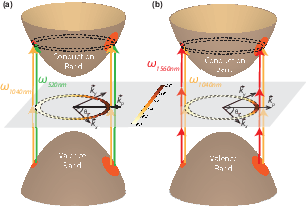}
\caption{\label{fig:conceptual}Conceptual diagrams of QuIC processes in semiconductors. In QuIC, the carriers injected into the conduction band are imbalanced in k-space. For the correct phase, the process is constructive for one direction in k-space, but destructive in the opposite direction. The (a) 1+2 QuIC and (b) 2+3 QuIC processes have the same transition energy. The calculated carrier distributions are projected to the virtual plane, shown as rings. The carrier distributions are maximum along the polarization direction ($\bm{k}_p$).}
\end{figure}

In semiconductors, QuIC can produce a current without an applied bias because the symmetry between optical injection rates of carriers with opposite momentum is broken, as illustrated in Fig.~\ref{fig:conceptual}. This realization of QuIC was demonstrated in gallium arsenide (GaAs) using the interference between one-photon and two-photon (1+2 QuIC) absorption (Fig.~\ref{fig:conceptual}(a))~\cite{Atanasov1996,Hache1997}. More recently, it was implemented using the interference between two-photon and three-photon absorption (2+3 QuIC) in aluminum gallium arsenide (AlGaAs) (Fig.~\ref{fig:conceptual}(b)~\cite{Wang2019}. Current-injecting QuIC requires an interference term that is odd under $\mathbf{k}\to -\mathbf{k}$, 
which in the dipole approximation arises from even-$L$ $\times$ odd-$L$ partial-wave mixing. 
Thus 1+2 and 2+3 QuIC are symmetry-allowed to inject a net current in the 
electric-dipole-dominated regime. Beyond controlling current in GaAs, 1+2 QuIC has been used to control charge~\cite{Fraser1999}, spin~\cite{Huebner2003,Stevens2003}, and vector currents~\cite{Sederberg2020} in GaAs, to inject currents in graphene~\cite{Rioux2011,Sun2010}, topological insulators~\cite{Bas2015} and transition metal dichalcogenides~\cite{Cui2015}, and to generate terahertz radiation from GaAs~\cite{Cote1999}, air~\cite{Cook2000}, and transition metal dichalcogenides~\cite{He2024,Xing2024}. 

 Here we show that 2+3 QuIC does not just break inversion symmetry, i.e., control if the injected current goes to the left or the right, but it actually determines the in-plane direction of the carrier injection with the polarizations of the incident two-color light, as predicted theoretically~\cite{Mahon2019}. The calculated distributions are shown in Figs.~\ref{fig:conceptual}(a) and~\ref{fig:conceptual}(b). For both 1+2 QuIC and 2+3 QuIC, the distribution is centered around the polarization direction. For 1+2 QuIC, the angular distribution is dominated by the cosine dependence arising from the cancellation between currents in opposite directions. In contrast, for 2+3 QuIC the angular spread is narrower, showing that the generation process itself is directional and produces a beam of carriers in a well-defined direction. The possibility of generating directed beams of carriers in a solid opens new avenues for studying the evolution of ballistic transport and provides an optical method that is sensitive to anisotropic band structure.

 The prior demonstration of 2+3 QuIC~\cite{Wang2019} did not, and could not, characterize the angular distribution of the injected photocurrent because it uses a single electrode pair and rotated only one polarization. The resulting angular scans conflate changes in the measured current amplitude due to the difference in polarizations of two wavelengths and due to the angle between the polarization and crystal axes with a rotation of the injected current direction with respect to the electrodes due to the polarization rotation. Moreover, a stationary current-injection pattern that rotates rigidly with the polarization axis requires joint (co-)rotation of the two polarizations, and therefore cannot be established by rotating only one color. The orthogonal-electrode-pair device used here, together with coordinated co-polarization rotation, resolves these ambiguities and enables measurement of the in-plane current vector in $k$-space.


Experimental demonstration of QuIC processes where the two photons have frequencies with an integer ratio, such as 1+2 QuIC, can be realized by using a single nonlinear harmonic generation step, e.g., second harmonic generation for 1+2 QuIC. However, for 2+3 QuIC, the ratio of the frequencies is not an integer, thus the correct frequencies cannot be achieved with a single harmonic generation step. The correct frequencies could be achieved by starting with a lower frequency performing harmonic generation of differing order on two copies of the lower frequency light. However, we choose to instead use frequency comb methods~\cite{Wang2019} where an initial frequency comb in spectrally broadened to span a frequency ratio of 3/2 and then the appropriate spectral regions are amplified. A matched comparison of 1+2 and 2+3 QuIC at the same total transition energy and in the same measurement geometry ensures that the narrower spread is a property of the higher-order interference process rather than a change in collection conditions.

\begin{figure}[htp]
\includegraphics[width=1\columnwidth]{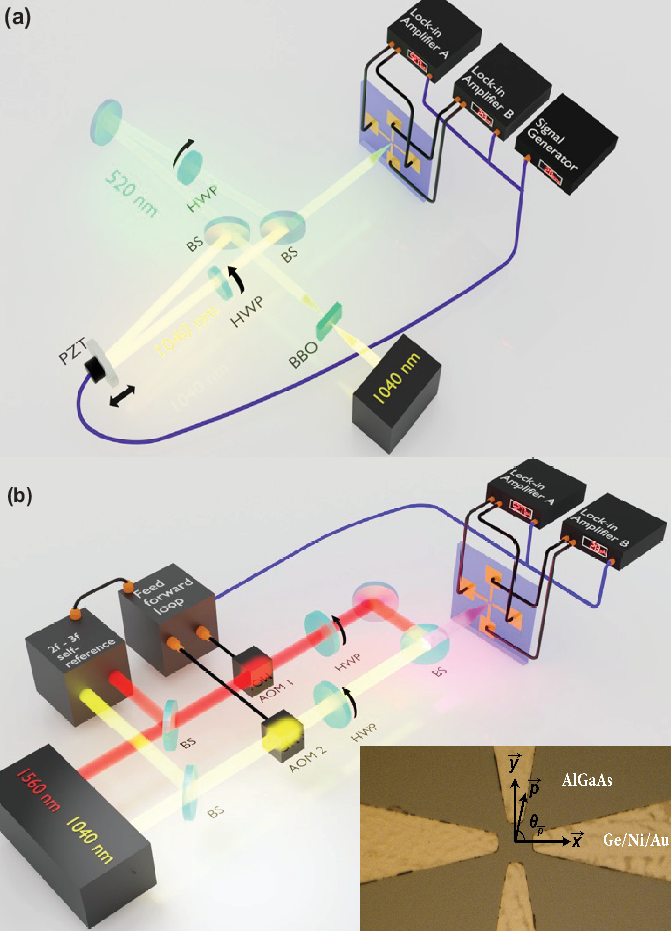}
\caption{\label{fig:apparatus}Conceptual experimental apparatus. (a) The experimental setup for 1+2 QuIC. (b) The experimental setup for 2+3 QuIC. (c) image of the sample showing the two sets of orthogonal electrodes. The $\bm{x}$ and $\bm{y}$ directions as well as the polarization direction, $\bm{p}$, are labeled.}
\end{figure}

As shown in Fig.~\ref{fig:apparatus}, we use two different setups to implement 1+2 QuIC and 2+3 QuIC. Both use the same light source, namely a modelocked erbium-doped fiber laser producing pulses spaced by 4 nanoseconds at a wavelength of 1560 nm that is amplified and spectrally broadened to reach a wavelength of 1040 nm, which is then amplified using a ytterbium-doped fiber amplifier at 1040 nm (2/3 the wavelength of the 1560 nm light). The light at 1560 and 1040 nm constitute sections of the same coherent frequency comb. The carrier-envelope phase of a frequency comb has been shown to determine the phase difference between different frequencies that determines if a QuIC process is constructive or destructive~\cite{Fortier2004} and control the direction of molecular disassociation~\cite{Kling2006}. We use the 1040 nm output, second harmonic generation, and a two-color interferometer to observe 1+2 QuIC as it is simpler than generating the octave spanning spectrum needed to observe 1+2 QuIC using frequency comb methods~\cite{Fortier2004}. Note that Fig.~\ref{fig:apparatus} shows beams splitters for simplicity; the actual setup uses a prism sequence to separate the colors. For 2+3 QuIC we use the evolving carrier-envelope phase of the coherent comb at 1560 and 1040 nm. Details of the setups are given in the supplementary material~\cite{SupplementalMaterial}. We verified that wavelength-dependent dispersion in the delivery optics does not significantly distort the durations of the 1560 nm and 1040 nm pulses at the sample: measured after the final delivery optics, both pulses are ~85 fs. Temporal overlap at the sample was set using a motorized delay stage and optimized by maximizing the QuIC current as a function of inter-pulse delay (supplementary Fig. S4), providing an experimental cross-correlation of the two pulse envelopes.
The 1+2 QuIC measurements used photon fluences of approximately \(3.1\times10^{15}\) and \(3.0\times10^{14}~\mathrm{photons/cm^2}\) per pulse for the \(1040~\mathrm{nm}\) and \(520~\mathrm{nm}\) beams, respectively. The 2+3 QuIC measurements used photon fluences of approximately \(5.9\times10^{15}\) and \(1.0\times10^{16}~\mathrm{photons/cm^2}\) per pulse for the \(1040~\mathrm{nm}\) and \(1560~\mathrm{nm}\) beams, respectively. These fluences are approximately a factor of 4 above the level needed for a signal-to-noise ratio of 1. The supplementary material provides further parameters \cite{SupplementalMaterial}.

To allow and select the desired order processes, a material with the correct bandgap energy must be used. For 1+2 QuIC, the light undergoing one-photon absorption must have a photon energy greater than the gap energy, while the light undergoing two-photon absorption must have energy below the gap. For 2+3 QuIC, both components of the incident light must be below the gap while the light undergoing two-photon absorption must have a photon energy above half-gap while the light undergoing three-photon absorption must have an energy below half-gap to suppress two-photon absorption but above one-third of the bandgap energy. To meet these requirements, we use AlGaAs because the bandgap energy can be tuned via the aluminum concentration to optimize the 2+3 QuIC process~\cite{Muniz2019}. Al$_x$Ga$_{1-x}$As enables bandgap engineering via the Al fraction $x$.  For the Al$_{0.28}$Ga$_{0.72}$As layer used here, the direct band gap corresponds to $E_g(x)\approx 1.77\,\mathrm{eV}$~\cite{Adachi_GaAs1994}. We fabricate electrodes with ohmic contacts to avoid complications due to field-induced QuIC~\cite{Wahlstrand2011}. The gap between the electrodes is 7 microns. Details are given in the supplementary material~\cite{SupplementalMaterial}.

\begin{figure}[htp]
\includegraphics[width=1.0\columnwidth]{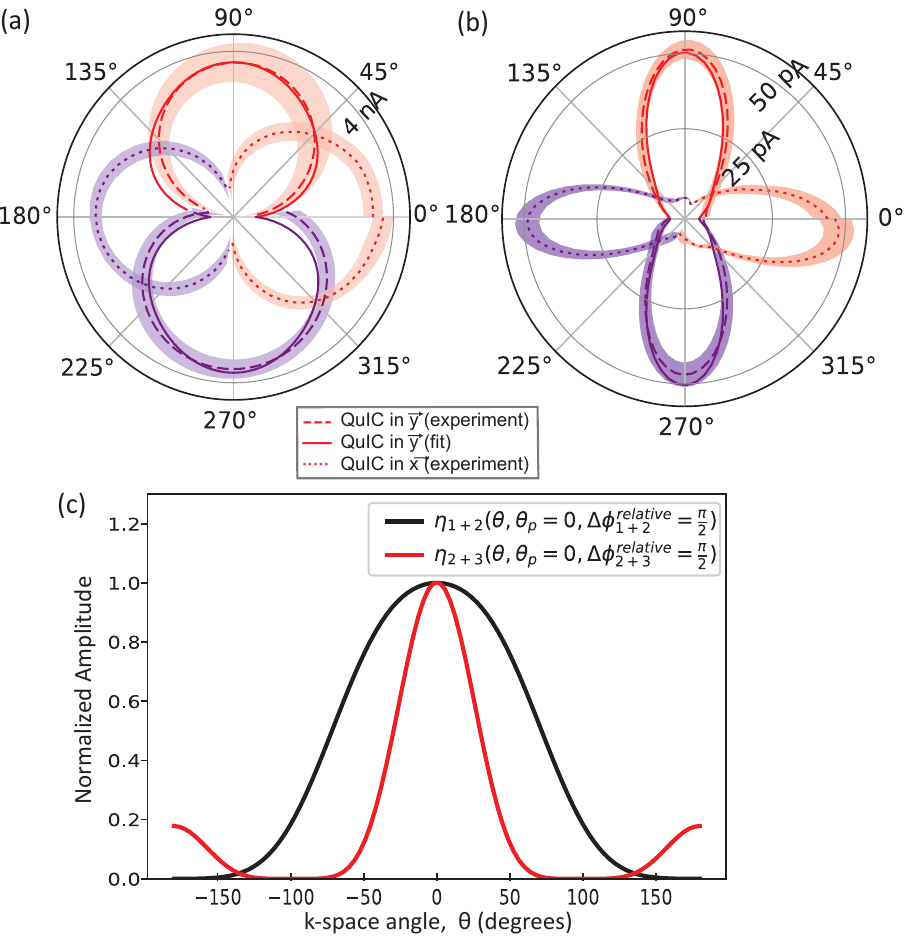}
\caption{\label{fig:copolarized} Co-linearly polarized results demonstrating a directional photocurrent. Polar plots of the photocurrent produced by (a) 1+2 QuIC and (b) by 2+3 QuIC as a function of the angle between the polarization direction and the x-direction, The dashed line is the experimental result for the electrodes in the x-direction while the dotted line is the result for the electrodes in the y-direction. The shaded area is the standard deviation of the 30 consecutive traces of polarization dependence. The solid line is theory for the y-oriented electrodes. Red indicates a positive signal and purple indicates a negative signal. (c) calculated QuIC current injection rate ($\eta$) as function of angle in the k-space for both 1+2 and 2+3 QuIC. $\theta_k$ is the angle between a direction in the k-space and the polarization of co-linearly-polarized two-color field. The relative phase between the two colors ($\Delta\phi$) is adjusted to be $\pi/2$ for both 1+2 QuIC and 2+3 QuIC.}
\end{figure}

To demonstrate the directionality of the photocurrent, we use two pairs of electrodes that are oriented in perpendicular directions, as shown in Fig.~\ref{fig:apparatus}(c). We measure the photocurrent simultaneously for both pairs of electrodes while rotating the polarization direction of both beams, i.e., the 1040 and 520 nm beams for 1+2 QuIC or the 1560 and 1040 nm beams for 2+3 QuIC. The carrier injection rates in $\mathbf{k}$-space for the QuIC processes are
\begin{multline}
\frac{d}{dt}J^a_{1+2} = \mathrm{Re}\sum_{bcd} \eta^{abcd}_{1+2}(\theta,\theta_p)\, e^{i\Delta\phi_{1+2}}\, \\
\times E^b_{1040}E^c_{1040}E^d_{520} + \mathrm{c.c.}-J^a_{1+2}/\tau,
\label{eq:eta12}
\end{multline}
and
\begin{multline}
\frac{d}{dt}J^a_{2+3} = \mathrm{Re}\sum_{bcdef} \eta^{abcdef}_{2+3}(\theta,\theta_p)\, e^{i\Delta\phi_{2+3}}\, \\
\times E^b_{1560}E^c_{1560}E^d_{1560} E^e_{1040}E^f_{1040} + \mathrm{c.c.}-J^a_{2+3}/\tau,
\label{eq:eta23}
\end{multline}
where $\eta_{1+2}$ and $\eta_{2+3}$ are the nonlinear response tensors that encode the interference between the two absorption pathways, $\theta$ is the angle of a wave-vector $\mathbf{k}$ in the sample plane, $\theta_p$ is the angle of the (co-rotated) linear polarization, and $\Delta\phi_{1+2}$, $\Delta\phi_{2+3}$ are the relative phases between the two colors---adjusted to $\pi/2$ in both experiments to maximize current injection. The $\eta$ tensors carry the directional information shown as the colored rings in Fig.~\ref{fig:conceptual}, their odd-in-$\mathbf{k}$ angular content produces a net current while the narrower angular spread of $\eta_{2+3}$ yields the directional ``beam'' that is the central result of this work. The measured photocurrent is shown in Fig.~\ref{fig:copolarized} as a function of the angle between the polarization direction and one set of electrodes, which we define as the x-direction corresponding to $\bm{k}_x$ in momentum space as shown in Fig.~\ref{fig:conceptual}, while the orthogonal set of electrodes is in the y-direction ($\bm{k}_y$). The dashed lines in Figs.~\ref{fig:copolarized}(a) and~\ref{fig:copolarized}(b) show the current collected on electrodes in the x-direction while the dotted line shows the current for the electrodes in the y-direction. The color indicates the relative signs of the signals, with red being positive and blue being negative. We choose to define the signal to be positive for the polarization oriented along the x-direction. Clearly in both cases, the current is maximized when the polarization is oriented along the axis of one of the electrode pairs. However, as the polarization direction is rotated, the current drops more rapidly for 2+3 QuIC as compared to 1+2 QuIC, as expected if the carriers have a narrower spread momentum spread in the plane of the sample, i.e., they are forming a beam in a well-defined direction.

To validate our measurements, we compare them to calculations of the angular dependence of the QuIC currents using the elements of the nonlinear susceptibility for AlGaAs following the derivation by Muniz et al.~\cite{Muniz2019}. We take into account the finite width of the electrodes by integrating over the angle that they intercept with respect to the focal spot. See the supplementary materials for the details of the calculation~\cite{SupplementalMaterial}. The results of the calculation for the y-oriented electrode are shown as a solid line in Figs.~\ref{fig:copolarized}(a) and~\ref{fig:copolarized}(b). The theory and experiment are in excellent agreement. To emphasize the narrower angular distribution of the current for 2+3 QuIC compared to 1+2 QuIC, we plot the calculated k-space current injection rate, without the integration over the finite electrodes, for both in Fig.~\ref{fig:copolarized}(c), showing that the 2+3 QuIC signal is narrower and clearly goes to zero, confirming that it is not producing carriers at all angles.

These results show that both the 1+2 QuIC and the 2+3 QuIC signals exhibit a one-fold symmetry with respect to the angle of the polarization of the incident light when both fields are linear polarized in the same direction and their angles are rotated together. Furthermore, the signals for the two electrode pairs are basically identical but simply offset by 90°. The theory used to validate these data predicts that rotating the polarization of one beam while the polarization of the other remains fixed will provide different results. Namely that if the field undergoing an odd-order interaction (1-photon or 3-photon absorption) is rotated, there will be dominantly a one-fold symmetry whereas if the field undergoing an even order interaction (2-photon absorption) is rotated there will be a two-fold symmetry. These results show that when the polarizations of the incident fields are not co-aligned, the QuIC process does not necessarily inject carriers in a well-defined direction.

\begin{figure}[htp]
\includegraphics[width=1.0\columnwidth]{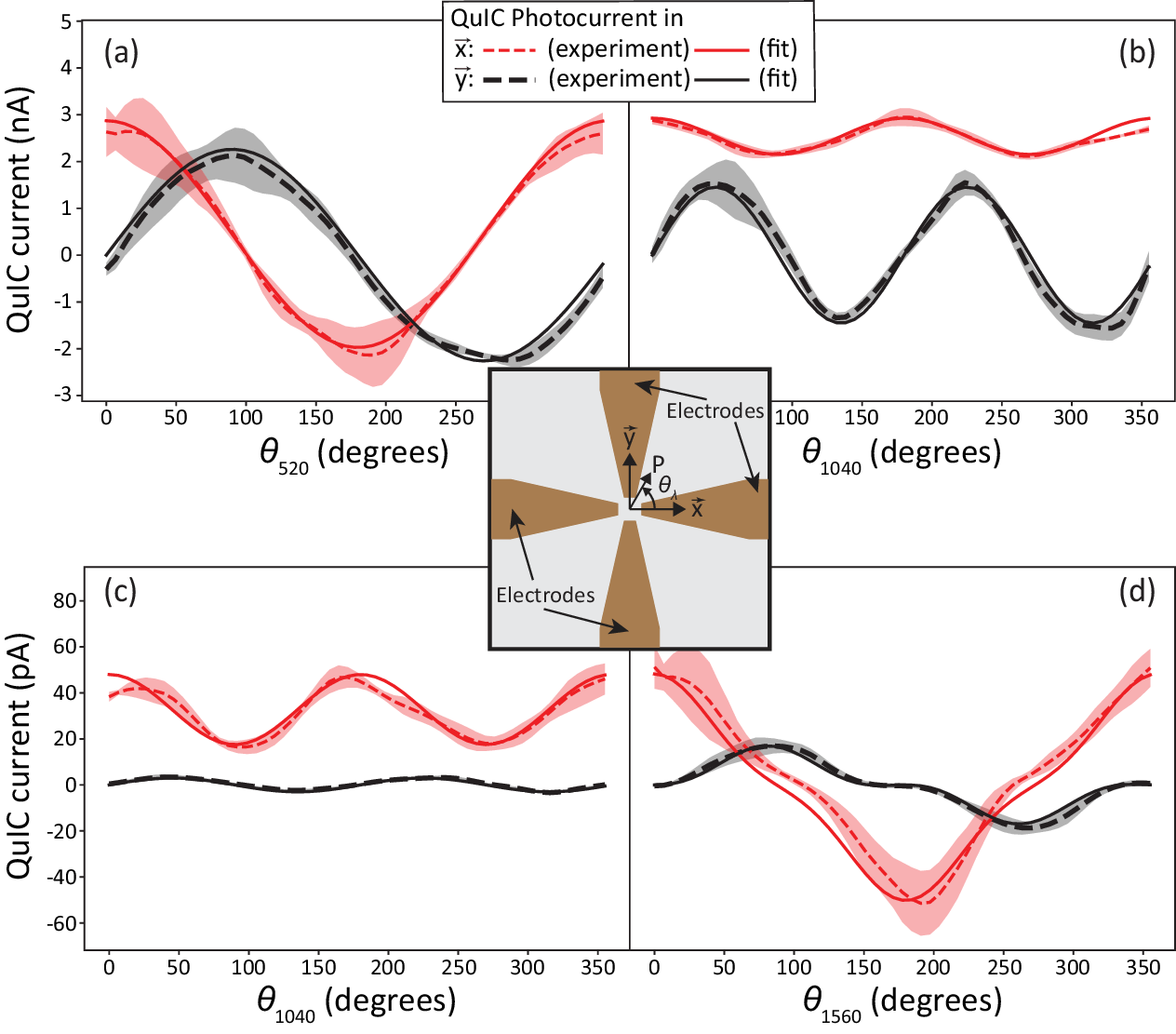}
\caption{\label{fig:noncopolarized}Non-colinear polarization dependencies. (a) \& (b) The non-colinear polarization dependencies of the 1+2 QuIC current and (c) \& (d) the 2+3 QuIC current. The inset shows a sketch of the electrodes and angle of the polarization being rotated, $\theta_\lambda$. The red dashed lines and black dashed lines are the average QuIC currents collected by the x-direction and y-direction electrodes, respectively. The shaded areas represent the measurement uncertainty. The solid lines are the theory results. In all plots the non-rotating polarization is in the x-direction. In (a), the polarization of 520 nm light is rotated. 1+2 QuIC has a one-fold symmetry with respect to the angle of the 520 nm polarization, $\theta_{520}$. In (b), the polarization of 1040 nm light is rotated. 1+2 QUIC has a two-fold symmetry with respect to the angle of the 1040 nm polarization, $\theta_{1040}$. In (c), the polarization of 1040 nm light is rotated. 2+3 QUIC has a two-fold symmetry with respect to the angle of the 1040 nm polarization, $\theta_{1040}$. In (d), the polarization of 1560 nm light is rotated. 2+3 QUIC has a one-fold symmetry with respect to the angle of the 1560 nm polarization, $\theta_{1560}$.}
\end{figure}

In Fig.~\ref{fig:noncopolarized}, we present the results of the corresponding experiments where the polarization direction of one of the four incident fields is rotated while the other is held fixed in the x-direction. The experimental results agree well with the theory, not just in the rotational symmetry discussed above, but also the amplitudes and offsets of the signal. The agreement further confirms and validates our experimental results and interpretation.

These results clearly show that the QuIC processes produce an electrical current in the direction of the polarization of the incident fields when they are co-linearly polarized. By using two sets of orthogonal electrodes, we exclude the possibility that the results are just due to a polarization dependent reduction in the QuIC current but rather must be due to the directionality of the current. The narrower angular distribution of the 2 + 3 QuIC signal as compared to the 1 + 2 QuIC signal shows that it is generated in a well-defined direction. Higher order processes will have even narrower angular distributions.

The sub-micron to micron scale distance over which we observe the QuIC current is much longer than the tens to hundreds of nanometers scale associated with ballistic transport in gallium arsenide and related compounds. However, the conditions are very different. Ballistic transport describes the evolution of the energy and momentum of electrons with a thermal energy distribution that are initially at rest prior to an electric field being switched on. In contrast, QuIC injects electrons with a transient highly non-thermal distribution, a large initial momentum, and no applied electric field. Due to the lack of an applied electric field, scattered carriers do not contribute to the QuIC current; it is solely due to ballistic carriers, whereas in a traditional transport measurement, scattered carriers contribute to a diffusive current because of the applied electric field. Thus, it is not surprising that the relevant length scales are significantly different. Indeed, our results provide a new avenue for testing and quantifying some of the underlying assumptions of transport theory. For example, transport models often assume energy dependent energy relaxation and momentum relaxation rates that are typically indirectly inferred from experimental measurements. Detailed studies of how our observations depend on sample temperature, geometry and material composition could provide a direct test of these assumptions and measurement of the relevant parameters.

Optical measurements are typically insensitive to the details of band structure, which usually need to be characterized using methods such as angle resolved photoemission spectroscopy to provide sensitivity to both the energy and direction in momentum space. Our results provide a path towards optical measurements that can probe specific regions in momentum space without needing to generate and analyze photoelectrons.

\begin{acknowledgments}
We acknowledge support from the National Science Foundation grant DMR-2004286. K.W. acknowledges support from the National Natural Science Foundation of China under grant 12174459.
\end{acknowledgments}

\bibliography{QuIC-arXiv}

\end{document}